\documentclass{article}

\usepackage{amsmath}
\usepackage{cite}

\newcommand{\ri}{\mathrm{i}}
\newcommand{\st}{\mathrm{st}}

\begin{document}
\begin{titlepage}
\begin{center}
{\large \textbf{Dynamical phase transition in one-dimensional
kinetic Ising model with nonuniform coupling constants}} \vskip
2\baselineskip \centerline {\sffamily Mohammad Khorrami
\footnote{e-mail: mamwad@mailaps.org} \&
 Amir Aghamohammadi\footnote
 {e-mail: mohamadi@alzahra.ac.ir}}
 \vskip 2\baselineskip
{\it Department of Physics, Alzahra University, Tehran 19384,
IRAN}
\end{center}
\vskip 2cm {\bf PACS numbers:} 64.60.-i, 05.40.-a, 02.50.Ga

\noindent{\bf Keywords:} reaction-diffusion, phase transition,
Glauber model

\begin{abstract}
\noindent An extension of the Kinetic Ising model with nonuniform
coupling constants on a one-dimensional lattice with boundaries is
investigated, and the relaxation of such a system towards its
equilibrium is studied. Using a transfer matrix method, it is
shown that there are cases where the system exhibits a
dynamical phase transition. There may be two phases, the fast phase
and the slow phase. For some region of the parameter space,
the relaxation time is independent of the reaction rates at the
boundaries. Changing continuously the reaction rates at the
boundaries, however, there is a point where the relaxation times
begins changing, as a continuous (nonconstant) function of
the reaction rates at the boundaries, so that at this point
there is a jump in the derivative of the relaxation time with
respect to the reaction rates at the boundaries.
\end{abstract}
\end{titlepage}
\newpage
\section{Introduction}
The Glauber dynamics was originally proposed to study the
relaxation of the Ising model towards equilibrium states \cite{RG}.
It is a simple non-equilibrium model of interacting spins with
spin-flip dynamics. Kinetic models based on the Ising model,
for example the Glauber model or the Kawasaki spin-pair exchange
mechanism model \cite{KK}, are phenomenological models and have
been extensively studied \cite{RG,KK,AF,TV,SSG,GL}. It has been
shown  that there is a relationship between the one-dimensional
kinetic Ising model at zero temperature and diffusion annihilation
in one dimension \cite{AF}. In \cite{TV}, using a damage spreading
method, the sensitivity of the time evolution of a kinetic Ising
model with Glauber dynamics against the initial conditions has
been investigated. The full time dependence of the space-dependent
magnetization and of the equal time spin-spin correlation functions
were studied in \cite{SSG}. Non-equilibrium two-time correlation
and response functions  for the ferromagnetic Ising chain with
Glauber dynamics have been studied in \cite{GL,LZ}.

Combinations of the Glauber and the Kawasaki dynamics have been
also considered \cite{DRS,GF,AT}. Most studies are focused on
uniform lattices where reaction rates are site-independent. It is
known that the ordinary Glauber model on a one-dimensional lattice
with boundaries at any temperature, shows a dynamical phase
transition\cite{MA}. The dynamical phase transition is controlled
by the rate of spin flip at the boundaries, and is a discontinuous
change of the derivative of the relaxation time towards the
stationary configuration. Among the simplest generalizations
beyond a uniform system is a lattice with alternating rates. In
\cite{SchSch02,SchSch03,MObZ}, the steady state configurational
probabilities of an Ising spin chain driven out of equilibrium by
a coupling to two heat baths have been investigated. An example is
a one-dimensional Ising model on a ring, in which the evolution is
according to a generalization of Glauber rates, such that spins at
even (odd) lattice sites experience a temperature $T_e$ ($T_o$).
In this model the detailed balance is violated. The response
function to an infinitesimal magnetic field for the Ising-Glauber
model with arbitrary exchange couplings has been studied in
\cite{Chatelain}. Other generalizations of the Glauber model
consist of, for example, alternating-isotopic chains and
alternating-bound chains (\cite{GO} for example). In \cite{MA1},
an asymmetric generalization of the zero-temperature Glauber model
on a lattice with boundaries was introduced. There it was shown
that in the thermodynamic limit (when the lattice becomes
infinite) the system shows two kinds of phase transitions. One of
these is a static phase transition, the other a dynamic one. The
static phase transition is controlled by the bulk reaction rates,
and is a discontinuous change of the behavior of the derivative of
the stationary magnetization at the end points, with respect to
the reaction rates. The dynamic phase transition is controlled by
the spin flip rates of the particles at the end points, and is a
discontinuous change of the relaxation time towards the stationary
configuration. Other phase transitions induced by boundary
conditions have also been studied (\cite{HS,RIK,AM2} for example).
Another generalization of the Glauber model was introduced in
\cite{SAM}. In this generalization, the processes are the same as
those of the ordinary Glauber model, but the rates depend on three
free parameters, rather than one free parameter as in the ordinary
Glauber model. Finally, this model was further generalized to the
case where the number of interacting sites is more than three and
the number of states at each site is more than two. This model too
violates detailed balance. In \cite{MDG}, an inherent spin
anisotropy (kinetic disorder) in the Glauber- Ising model was
introduced. It was shown both analytically and numerically, that
there is  a slow logarithmic factor in the decay of the density of
kinks at large times. In \cite{MO}, the problem of the effect of
quenched impurities on one-dimensional non-equilibrium Glauber-
Ising-type models has been investigated numerically. It was shown
that the model has a continuous phase transition to an absorbing
state. Also a mean-field approach has been used to study the
Glauber-type stochastic dynamics of a  model on a square lattice
in which two interpenetrating square sublattices have spins that
can take two values, alternated with spins that can take the four
values \cite{DKC}.

The behavior of an Ising model with nonuniform coupling constants
at low temperatures on a one dimensional periodic lattice has been
discussed in \cite{AR}. The static behavior of an Ising model with
nonuniform coupling constants on a one-dimensional lattice with
boundaries was investigated in \cite{MA2}. Detailed balance was used
in \cite{AR,MA2} to propose reaction rates for the system. In \cite{MA2}
the time-independent solution to the evolution equation of the
expectation values of spins was studied. This solution was expressed
in terms of a transfer matrix. While it is true that the ordinary
Ising model does not exhibit any phase transition in finite temperatures
(the expectation values of the spins vanish if there is no external
magnetic field), this is not necessarily the case for the Ising
model with inhomogeneous boundary conditions. It was shown that in
the thermodynamic limit different phases could occur for this
system, according to whether the eigenvalues of the transfer
matrix are less than or larger than one.

In this paper the dynamical properties of kinetic Ising model with
nonuniform coupling constants are investigated. The nonuniformity
could arise from either a nonuniformity in the coupling constants or a
nonuniformity in the temperature. However, as long as the dynamics of
the system is considered, only the ratios of the coupling constants
and the temperature are important, so without loss of generality one
can assume the temperature to be constant and put all the nonuniformity
in the coupling constant. So a nonuniformity in the (effective) coupling
constants could be due to a nonuniformity in the temperature. It could also
be due to a nonuniformity in the interaction between neighboring spins, or
due to a nonuniformity in some sort of inertia, which opposes the changes
in the state of the system.

It is shown that there are cases where the system exhibits a dynamical phase
transition. There are two phases: the fast phase and the slow
phase. For some region of the parameter space, the relaxation time
is independent of the reaction rates at the boundaries. This is
the fast phase. Changing continuously the reaction rates at the
boundaries, there is a point where the relaxation time begins
increasing, so that at this point there is a jump in the
derivative of the relaxation time with respect to the reaction
rates at boundaries. This is the dynamical phase transition and
the region where the relaxation time changes with reaction rates
at boundaries is the slow phase. So the dynamical phase transition
studied here, is a discontinuity in the derivative of the relaxation
time (from zero to nonzero) with respect to reaction rates at
the boundaries.

The scheme of the paper is as follows. In section 2, the model is
introduced, and the evolution equation for the spin expectation
values is obtained. In section 3 the relaxation of the spin
expectation values towards their stationary state is investigated,
through a study of the time scales of this relaxation, and a
general equation is obtained for these time scales. In section 4,
some examples are studied in more detail. It is seen that some of
these show two phases, the fast phase and the slow phase. Section
5 is devoted to the concluding remarks, and section 6 is an appendix
on the site-link notation and the relaxation time.

\section{One-dimensional Ising model with nonuniform coupling constants}
Consider a one-dimensional lattice with $(L+1)$ sites, labeled
from $0$ to $L$. At each site $i$, there is a spin variable $s_i$,
which could be $+1$ (spin up, $\uparrow$), or $-1$ (spin down,
$\downarrow$). These spins interact through the Hamiltonian,
\begin{equation}\label{dh.01}
{\mathcal H} ={\mathcal H}'_0+\left(\sum_\alpha{\mathcal
H}_\alpha\right)+{\mathcal H}'_L,
\end{equation}
where ${\mathcal H}_\alpha$ is the Hamiltonian interaction for the
link $\alpha$:
\begin{equation}\label{dh.02}
{\mathcal H}_\alpha =-J_\alpha\,s_{\alpha-\mu}\, s_{\alpha+\mu},
\end{equation}
$J_\alpha$ is the coupling constant in the link $\alpha$, and
\begin{equation}\label{dh.03}
\mu:=\frac{1}{2}.
\end{equation}
The link $\alpha$ links the sites $\alpha-\mu$ and $\alpha+\mu$,
so that $\alpha\pm\mu$ are integers, and $\alpha$ runs from $\mu$
up to $(L-\mu)$. Throughout this paper, sites are denoted by Latin
letters which represent integers, while links are denoted by Greek
letters which represent integers plus one half ($\mu$), so that
the link $\alpha$ joins the sites $(\alpha-\mu)$ and
$(\alpha+\mu)$, while the site $i$ joins the links $(i-\mu)$ and
$(i+\mu)$. ${\mathcal H}'_0$ and ${\mathcal H}'_L$ correspond to
interactions at the boundaries. A more extended explanation of the
site-link notation is presented in the appendix.

A nonuniform Glauber model gives the dynamics of the Ising model
with nonuniform coupling constants ($J_\alpha$) such that the rate
of a spin flip is determined through its interaction with its two
neighboring sites and a heat bath at the temperature $T$.
Introducing
\begin{equation}\label{dh.04}
K_\alpha:=\frac{J_\alpha}{k_\mathrm{B}\,T},
\end{equation}
where $k_\mathrm{B}$ is the Boltzmann's constant, it was shown in
\cite{MA2} that assuming nearest neighbor interaction, and that in
each step only one spin flips, detailed balance gives
\begin{equation}\label{dh.05}
\omega[(S', s_j)\to(S',-s_j)]=\Gamma_j\,[1-s_j\,
\tanh(K_{j-\mu}\,s_{j-1}+K_{j+\mu}\,s_{j+1})],
\end{equation}
where $S'$ denotes the configuration of the lattice apart from
site $j$, and $\Gamma_j$'s are constants. So that the spin at the
site $j$ flips according to the following interactions with the
indicated rates.
\begin{align}\label{dh.06}
\uparrow \; \uparrow \; \uparrow \; \to \; \uparrow \;
\downarrow\; \uparrow \mbox{\quad and\quad} \downarrow \;
\downarrow \; \downarrow\; \to \; \downarrow \; \uparrow \;
\downarrow &\mbox{\quad with rate\quad} 1-\tanh(K_{j-\mu}+K_{j+\mu}), \nonumber \\
\uparrow \; \downarrow \; \uparrow \; \to \; \uparrow \;
\uparrow\; \uparrow \mbox{\quad and\quad} \downarrow \; \uparrow
\; \downarrow\; \to \; \downarrow \; \downarrow \;
\downarrow &\mbox{\quad with rate\quad} 1+\tanh(K_{j-\mu}+K_{j+\mu}), \nonumber \\
\uparrow \; \uparrow \; \downarrow \; \to  \; \uparrow \;
\downarrow \; \downarrow \mbox{\quad and\quad} \downarrow\;
\downarrow \; \uparrow \; \to \;\downarrow \; \uparrow \;
\uparrow  & \mbox{\quad with rate\quad} 1-\tanh(K_{j-\mu}-K_{j+\mu}),\nonumber \\
\downarrow \; \uparrow \; \uparrow \; \to \; \downarrow \;
\downarrow \; \uparrow \mbox{\quad and\quad} \uparrow \;
\downarrow \; \downarrow \; \to  \; \uparrow \; \uparrow \;
\downarrow  &\mbox{\quad with rate\quad}
1+\tanh(K_{j-\mu}-K_{j+\mu}),\qquad
\end{align}
where $\Gamma_j$'s have been taken independent of $j$, and set to
one by rescaling the time, as in \cite{AR,MA2}. At the boundaries,
there are other interactions as well (corresponding to $\mathcal{H}'_0$
and $\mathcal{H}'_L$). The spin of the zeroth site may flip like
\begin{align}\label{dh.07}
\uparrow \; \downarrow \;\to \;  \downarrow\;\downarrow \;
&\mbox{\quad with rate\quad} g_1, \nonumber \\
\uparrow \; \uparrow \; \to \;  \downarrow\;\uparrow \;
&\mbox{\quad with rate\quad} g_2, \nonumber\\
\downarrow \;\uparrow \; \to \;  \uparrow \; \uparrow \;
&\mbox{\quad with rate\quad} g_3, \nonumber \\
\downarrow\;\downarrow \; \to \;  \uparrow \;\downarrow\;
&\mbox{\quad with rate\quad} g_4,
\end{align}
and the spin of the $L$'th site may flip like
\begin{align}\label{dh.08}
\downarrow \;\uparrow \;\to \;  \downarrow\;\downarrow \;
&\mbox{\quad with rate\quad} h_1, \nonumber \\
\uparrow \; \uparrow \; \to \;  \uparrow \;\downarrow\;
&\mbox{\quad with rate\quad} h_2, \nonumber\\
\uparrow \;\downarrow \; \to \;  \uparrow \; \uparrow \;
&\mbox{\quad with rate\quad} h_3, \nonumber \\
\downarrow\;\downarrow \; \to \;  \downarrow\;\uparrow \;
&\mbox{\quad with rate\quad} h_4.
\end{align}
It is known that with the rates (\ref{dh.06}), the time derivatives
of the one-point functions in the bulk are expressed in terms of only the
one-point functions. To make this true for the boundaries
as well, the following relations should hold \cite{MA1}.
\begin{align}\label{dh.09}
g_1+g_4&=g_2+g_3, \nonumber \\
h_1+h_4&=h_2+h_3.
\end{align}
The first relation, for example, means that the sum of conditional spin flip
rates at the zeroth site is independent of the state of the first site. This
does not, however, means that the flip rate at the zeroth site is independent
of the state of the first site; since, for example, $(g_1+g_4)$ itself is not a
(conditional) spin flip rate.

Thus the evolution equation for the expectation values of the spins are
\begin{align}\label{dh.10}
\langle\dot s_j\rangle=\;&-2\,\langle s_j\rangle+
[\tanh(K_{j-\mu}+K_{j+\mu})+\tanh(K_{j-\mu}-K_{j+\mu})]\,\langle
s_{j-1}\rangle\nonumber \\&  +
[\tanh(K_{j-\mu}+K_{j+\mu})-\tanh(K_{j-\mu}-K_{j+\mu})]\, \langle
s_{j+1}\rangle, \quad 0<j<L\nonumber\\
\langle\dot s_0\rangle=\;&(g_3-g_1)-(g_2+g_3)\,\langle
s_0\rangle+(g_1-g_2)\,\langle s_1\rangle,\nonumber\\
\langle\dot s_L\rangle=\;&(h_3-h_1)-(h_2+h_3)\,\langle
s_L\rangle+(h_1-h_2)\,\langle s_{L-1}\rangle.
\end{align}
The static solution ($\langle s\rangle_{\st}$) was studied in
\cite{MA2}. Here the dynamics (the relaxation towards the static
solution) is addressed.

\section{The relaxation of the system towards equilibrium}
The homogeneous part of eq. (\ref{dh.10}) can be written as
\begin{equation}\label{dh.11}
\langle\dot s_j\rangle=h_j^l \,\langle s_l\rangle.
\end{equation}
A brief introduction regarding the relaxation time of the system has been contained
in the appendix. Denoting an eigenvalue of $h$ by $E$, and the corresponding eigenvector by
$x$, it is seen that
\begin{align}\label{dh.12}
E x_j=\;&-2\,x_j+
[\tanh(K_{j-\mu}+K_{j+\mu})+\tanh(K_{j-\mu}-K_{j+\mu})]\,x_{j-1}\nonumber
\\&  + [\tanh(K_{j-\mu}+K_{j+\mu})-\tanh(K_{j-\mu}-K_{j+\mu})]\,
x_{j+1}, \quad 0<j<L\\ \label{dh.13} E
x_0=\;&-(g_2+g_3)\,x_0+(g_1-g_2)\,x_1,\\ \label{dh.14}E
x_L=\;&-(h_2+h_3)\, x_L+(h_1-h_2)\,x_{L-1},
\end{align}
which  can be written as
\begin{equation}\label{dh.15}
X_{j+\mu}=\tilde D_j\,X_{j-\mu},
\end{equation}
where
\begin{equation}\label{dh.16}
X_\alpha:=\begin{pmatrix}x_{\alpha-\mu}\\  \\
x_{\alpha+\mu}\end{pmatrix},
\end{equation}
and
\begin{equation}\label{dh.17}
\tilde D_j:=\begin{pmatrix}0&1 \\ & \\
\displaystyle{-\frac{\sinh(2\,K_{j-\mu})} {\sinh(2\,K_{j+h})}}&
\displaystyle{\frac{\cosh(2\,K_{j-\mu})+\cosh(2\,K_{j+\mu})}
{\sinh(2\,K_{j+\mu})}}\,\left(1+\frac{E}{2}\right)\end{pmatrix}.
\end{equation}
Using the recursion relation (\ref{dh.15}) repeatedly, one arrives at
\begin{equation}\label{dh.18}
X_\alpha=\tilde D_{\alpha\,\beta}\,X_\beta,
\end{equation}
where
\begin{equation}\label{dh.19}
\tilde D_{\alpha\,\beta}:=\tilde D_{\alpha-\mu}\cdots \tilde D_{\beta+\mu}
\end{equation}
(\ref{dh.13}) and (\ref{dh.14}) are boundary conditions for $X$, and can be rewritten as
\begin{equation}\label{dh.20}
\begin{pmatrix}x_0\\  \\
x_1\end{pmatrix}\propto \begin{pmatrix}g_1-g_2\\ \\
g_2+g_3+E\end{pmatrix},
\end{equation}
and
\begin{equation}\label{dh.21}
\begin{pmatrix}x_{L-1}\\  \\
x_L\end{pmatrix}\propto \begin{pmatrix}h_2+h_3+E\\
\\ h_1-h_2\end{pmatrix},
\end{equation}
or
\begin{equation}\label{dh.22}
X_\mu\propto \begin{pmatrix}g_1-g_2\\ \\
g_2+g_3+E\end{pmatrix},
\end{equation}
and
\begin{equation}\label{dh.23}
X_{L-\mu}\propto \begin{pmatrix}h_2+h_3+E\\
\\ h_1-h_2\end{pmatrix}.
\end{equation}
Defining
\begin{equation}\label{dh.24}
W:=\begin{pmatrix}h_2-h_1&h_2+h_3+E\end{pmatrix},
\end{equation}
it is seen that
\begin{equation}\label{dh.25}
W\,X_{L-\mu}=0.
\end{equation}
Then, using (\ref{dh.18}) for $\alpha=L-\mu$, and $\beta=\mu$, one arrives at
\begin{equation}\label{dh.26}
W\,\tilde D_{L-\mu\,\mu}\,V=0,
\end{equation}
where
\begin{equation}\label{dh.27}
V:=\begin{pmatrix}g_1-g_2\\ \\
g_2+g_3+E\end{pmatrix}.
\end{equation}
(\ref{dh.26}) is a polynomial equation of order $(L+1)$ for $E$, which has
$(L+1)$ roots for $E$. All roots should have nonpositive real
parts. The root with largest real part determines the relaxation
time of the system. The aim is to investigate the behavior of this
root in the thermodynamic limit ($L\to\infty$).

\section{Special cases}
\begin{itemize}
\item[\textbf{1}] Uniform coupling constant:
\begin{equation}\label{dh.28}
K_\alpha=K.
\end{equation}
This is the same kinetic Ising model, the dynamical phase
transition of which induced by boundary terms was investigated in
\cite{MA1}, using a different method. To fix notation and become
familiar with the transfer matrix method, let's use this method
for this case. In this case $\tilde D_j$ is independent of $j$:
\begin{equation}\label{dh.29}
\tilde D:=\begin{pmatrix}0&1 \\ & \\ -1&
(2+E)\coth(2K)\end{pmatrix}.
\end{equation}
Denoting an eigenvalue of the the matrix $\tilde D$ by $z$, it is
seen that
\begin{equation}\label{dh.30}
E=-2+(z+z^{-1})\,\tanh(2K).
\end{equation}
So the product of the eigenvalues is one. (\ref{dh.26}) recasts to
\begin{equation}\label{dh.31}
W\,\tilde D^{L-1}\,V=0.
\end{equation}
$\tilde D$ can be written as
\begin{equation}\label{dh.32}
\tilde D=\frac{1}{z^{-1}-z}\,\begin{pmatrix}1&1 \\ & \\
z& z^{-1}\end{pmatrix}\,\begin{pmatrix}z&0 \\ & \\
0& z^{-1}\end{pmatrix}\,\begin{pmatrix}z^{-1}&-1 \\ & \\
-z& 1\end{pmatrix},
\end{equation}
from which one obtains
\begin{equation}\label{dh.33}
\tilde D^{L-1}=\frac{1}{z^{-1}-z}\,\begin{pmatrix}
z^{L-2}-z^{-L+2}&-z^{L-1}+z^{-L+1} \\ & \\
z^{L-1}-z^{-L+1}& -z^L+z^{-L}\end{pmatrix}.
\end{equation}
So (\ref{dh.31}) becomes
\begin{equation}\label{dh.34}
W\,\begin{pmatrix}
z^{L-2}-z^{-L+2}&-z^{L-1}+z^{-L+1} \\ & \\
z^{L-1}-z^{-L+1}& -z^L+z^{-L}\end{pmatrix}\,V=0.
\end{equation}
Substituting $E$, using (\ref{dh.30}), one arrives from the above
equation at a polynomial equation of order $(2\,L+4)$ for $z$. Two
roots of this equation are $\pm 1$, which are not the roots of
(\ref{dh.31}), as (\ref{dh.34}) has been obtained from
(\ref{dh.31}) by multiplying it by $(z^{-1}-z)$. The other
$(2\,L+2)$ roots consist of $(L+1)$ pairs. In each pair the two
roots are inverses of each other, and both roots of each pair give
the same value for $E$.

The roots of (\ref{dh.34}) for $z$ are either unimodular or not.
For a unimodular solution
\begin{equation}\label{dh.35}
z=:e^{\ri\,\theta},
\end{equation}
equation (\ref{dh.34}) becomes
\begin{equation}\label{dh.36}
W\,\begin{pmatrix}
\sin(L-2)\,\theta&-\sin(L-1)\,\theta \\ & \\
\sin(L-1)\,\theta& -\sin L\,\theta\end{pmatrix}\,V=0,
\end{equation}
which can be written like
\begin{equation}\label{dh.37}
F(\theta)=0.
\end{equation}
It is seen that in the thermodynamic limit, the entries of the
matrix appearing in the left-hand side of (\ref{dh.36}) change
sign when the value of $\theta$ is changed by $(\pi/L)$, while $V$
and $W$ do not change. So the sign of $F(\theta)$ is opposite of
the sign of $F[\theta+(\pi/L)]$, meaning that for any value of
$\theta_0$ there is at least one root for $\theta$ between
$\theta_0$ and $[\theta_0+(\pi/L)]$. So, in the thermodynamic
limit all of unimodular $z$'s are roots of (\ref{dh.34}). These
correspond to real values of $E$, with the maximum
\begin{equation}\label{dh.38}
E_0=-2+2\,\tanh (2\,|K|),
\end{equation}
which corresponds to the relaxation time
\begin{equation}\label{dh.39}
\tau_0=\frac{1}{2-2\tanh(2\,|K|)}.
\end{equation}
If there are no nonunimodular solutions for $z$, this is the
relaxation time of the system, which does not depend on the
reaction rates at the boundaries. Let's call this the fast phase.
If, however, there is a value for $E$ the real part of which is
larger than the right-hand side of (\ref{dh.38}), then the
relaxation time is larger than $\tau_0$ and, as it will be seen,
does depend on the boundaries. This is called the slow phase.

If there is a nonunimodualr root for $z$, then there is a root $z$
with modulus larger than one. Then, in the thermodynamic limit
(\ref{dh.34}) becomes (for this root)
\begin{equation}\label{dh.40}
W\,\begin{pmatrix}z^{-1}&-1 \\ & \\
1& -z\end{pmatrix}\,V=0,
\end{equation}
or
\begin{equation}\label{dh.41}
[(h_2-h_1)\,z^{-1}+h_2+h_3+E]\,[(g_2-g_1)\,z^{-1}+g_2+g_3+E]=0.
\end{equation}
Using (\ref{dh.30}), this equation turns out to be
\begin{align}\label{dh.42}
&\{[h_2-h_1+\tanh(2\,K)]\,z^{-1}+[\tanh(2\,K)]\,z+h_2+h_3-2\}\nonumber\\
\times&\{[g_2-g_1+\tanh(2\,K)]\,z^{-1}+[\tanh(2\,K)]\,z+g_2+g_3-2\}=0,
\end{align}
which is exactly the same as equation (27) in \cite{MA1}. So the
rest of discussion is exactly similar to that of \cite{MA1}. The
root obtained from (\ref{dh.42}) does depend on the reaction rates
at the boundaries. Such a root corresponds to a relaxation time
larger than $\tau_0$, iff
\begin{equation}\label{dh.43}
[\mathrm{Re}(z+z^{-1})]\,[\mathrm{sgn}(K)]>2.
\end{equation}
This gives the following coexistence surface for the two phases
\begin{equation}\label{dh.44}
g_2+g_3+[\mathrm{sgn}(K)]\,(g_2-g_1)-2\,[1-\tanh(2\,|K|)]=0,
\end{equation}
or
\begin{align}\label{dh.45}
&[4\,\tanh(2\,|K|)+2-g_2-g_3]\,[\tanh(2\,|K|)-g_1+g_2]\nonumber\\
+&[\tanh(2\,|K|)]\,(g_2+g_3-2)=0,
\end{align}
and similar solutions with $h_i$'s substituting $g_i$'s.

In summary, for some values of reaction rates at the boundaries
the relaxation time is $\tau_0$, which is independent of the
reaction rates at the boundaries. At a certain point, however, the
relaxation time starts increasing from $\tau_0$ and being
dependent on the reaction rates at the boundaries. The dynamical
phase transition is this discontinuous change of the derivative of
the relaxation time towards the stationary configuration.

As examples for this transition, let's consider two cases. Regarding
the coexistence curve (\ref{dh.44}), it is seen that for
\begin{align}\label{dh.46}
g_1&=g_2,\nonumber\\
g_4&=g_3,
\end{align}
one has
\begin{equation}\label{dh.47}
\begin{cases}g_2+g_3<2[1-\tanh(2\,|K|)],&\hbox{slow phase}\\
g_2+g_3>2[1-\tanh(2\,|K|)],&\hbox{fast phase}
\end{cases}.
\end{equation}
Regarding the coexistence curve (\ref{dh.45}), it is seen that for
\begin{align}\label{dh.48}
g_1&=g_3,\nonumber\\
g_4&=g_2,\nonumber\\
g_2+g_3&=2,
\end{align}
one has
\begin{equation}\label{dh.49}
\begin{cases}g_3-g_2<\tanh(2\,|K|),&\hbox{slow phase}\\
g_3-g_2>\tanh(2\,|K|),&\hbox{fast phase}
\end{cases}.
\end{equation}

\item[\textbf{2}] $K_\alpha=\begin{cases}K',& \alpha<L'\\ K,& \alpha>L'\end{cases}$

It has been assumed that in the thermodynamic limit both $L'$ and
$(L-L')$ tend to infinity. Also, without loss of generality it can
be assumed that $|K|>|K'|$. This is a lattice consisting of two
parts, on each part the coupling constant is uniform.
Equation (\ref{dh.26}) becomes
\begin{equation}\label{dh.50}
W\,\tilde D^{L-L'-1}\,\tilde D_{L'}\,\tilde D'^{L'-1}\,V=0,
\end{equation}
where $\tilde D$ is defined through (\ref{dh.29}), $\tilde D'$ is
similar to $\tilde D$ but with $K$ replaced by $K'$, and
\begin{equation}\label{dh.51}
\tilde D_{L'}:=\begin{pmatrix}0&1 \\ & \\
\displaystyle{-\frac{\sinh(2\,K')} {\sinh(2\,K)}}&
\displaystyle{\frac{\cosh(2\,K')+\cosh(2\,K)}
{\sinh(2\,K)}\,\left(1+\frac{E}{2}\right)}\end{pmatrix}.
\end{equation}
Denoting an eigenvalue of $\tilde D$ ($\tilde D'$) by $z$ ($z'$),
it is seen that (\ref{dh.30}) holds and
\begin{equation}\label{dh.52}
E=-2+(z'+z'^{-1})\,\tanh(2K').
\end{equation}
Equations (\ref{dh.50}), (\ref{dh.30}), and (\ref{dh.52}) are used
to obtain the values of $z$ and $z'$. First assume that $z$ is
unimodular. Then $E$ is real and from (\ref{dh.52}) it is seen
that $z'$ is either unimodular or real. If $z'$ is unimodular too,
then the possible values for $z$ are not all of the points of
the unit circle. Defining $\theta$ by (\ref{dh.35}), it is seen
that those values of $\theta$ which correspond to unimodular
values for $z'$ are
\begin{equation}\label{dh.53}
\cos^{-1}\left[\frac{\tanh(2\,|K'|)}{\tanh(2\,|K|)}\right]\leq|\theta|
\leq\pi-
\cos^{-1}\left[\frac{\tanh(2\,|K'|)}{\tanh(2\,|K|)}\right].
\end{equation}
For each value of $\theta$ out of this region, there are two real
values for $z'$ which are inverses of each other. For that value
of $z'$ the modulus of which is greater than one, one arrives from
(\ref{dh.50}) at
\begin{equation}\label{dh.54}
W\,\begin{pmatrix}
\sin(L-L'-2)\,\theta&-\sin(L-L'-1)\,\theta\\ & \\
\sin(L-L'-1)\,\theta& -\sin(L-L')\,\theta\end{pmatrix}\,\tilde D_{L'}\,\begin{pmatrix}z'^{-1}&-1 \\ & \\
1& -z'\end{pmatrix}\,V=0,
\end{equation}
again showing that if $(L-L')$ tends to infinity, all of the
(remaining) values of $\theta$ are roots of this equation. So one
piece of the solutions for $E$ corresponds to the full unit circle
for $z$ and the full unit circle plus two real segments for $z'$.
The relaxation time corresponding to these eigenvalues is $\tau_0$
from (\ref{dh.39}). It is seen that it is the larger coupling
which determines the relaxation time (provided the part of the
lattice which corresponds to this coupling becomes infinite). If
all of the solutions for $z$ are unimodular, $\tau_0$ is the
relaxation time of the system and the system is in the fast phase.

The system would be in the slow phase (where the relaxation time
is larger than $\tau_0$ and does depend on the rates at the
boundaries) iff there exists a solution for $z$ for which
(\ref{dh.43}) holds. In this case, one arrives at
\begin{equation}\label{dh.55}
W\,\begin{pmatrix}z^{-1}&-1 \\ & \\
1& -z\end{pmatrix}\,\tilde D_{L'}\,\begin{pmatrix}z'^{-1}&-1 \\ & \\
1& -z'\end{pmatrix}\,V=0,
\end{equation}
where $z$ and $z'$ have are solutions which have moduli larger
than one. One then arrives at
\begin{align}\label{dh.56}
&\{[h_2-h_1+\tanh(2\,K)]\,z^{-1}+[\tanh(2\,K)]\,z+h_2+h_3-2\}\nonumber\\
\times&[z\,\sinh(2\,K)+z'\,\sinh(2\,K')]\nonumber\\
\times&\{[g_2-g_1+\tanh(2\,K')]\,z'^{-1}+[\tanh(2\,K')]\,z'+g_2+g_3-2\}=0.
\end{align}
From (\ref{dh.30}) and (\ref{dh.52}), it is seen that the real
parts of $[z\,\tanh(2\,K)]$ and $[z'\,\tanh(2\,K')]$, and hence
the real parts of $[z\,\sinh(2\,K)]$ and $[z'\,\sinh(2\,K')]$ have
the same sign, or both are zero. So the second factor in the left
hand side of (\ref{dh.56}) does not vanish unless $z$ and $z'$ are
$(\pm\ri)$. But these values of $z$ and $z'$ are unimodular hence
not acceptable, as the roots of (\ref{dh.56}). So in order that
(\ref{dh.56}) is satisfied, either the first or the third factor
in the left hand side of (\ref{dh.56}) should vanish. This is
similar to (\ref{dh.42}), with the couplings corresponding to each
part of the lattice instead of the uniform coupling. One then
arrives at (\ref{dh.44}) and (\ref{dh.45}) with $K$ replaced by
$K'$, or similar expressions with $K'$ and $g_i$'s replaced by $K$
and $h_i$'s respectively, as the coexistence surface. Of course
from the possible roots of (\ref{dh.56}), that root is chosen
which corresponds to the larger relaxation time (or larger real
part for $E$), and one should also check that this root satisfies
(\ref{dh.43}).

To summarize, in the fast phase the relaxation time is determined
by the coupling with larger modulus. In the slow phase the two
ends of the lattice behave independently, the relaxation time
being determined by the largest.

\item[\textbf{3}] $K_\alpha=\begin{cases}K',&\alpha<L'\\
K_\alpha,&L'<\alpha<L-L''\\ K'',&L-L''<\alpha\end{cases}$

It has been assumed that in the thermodynamic limit both $L'$ and
$L''$ tend to infinity, while $(L-L'-L'')$ remains finite.
This is a lattice with a finite part inside which, where outside
that part the coupling is uniform. Equation (\ref{dh.26}) becomes
\begin{equation}\label{dh.57}
W\,\tilde D''^{L''-1}\,\tilde D_{L-L''+\mu\;L'-\mu}\,\tilde
D'^{L'-1}\,V=0,
\end{equation}
where $\tilde D'$ and $\tilde D''$ are defined through
(\ref{dh.29}), but with $K$ replaced by $K'$ and $K''$
respectively. Similar to the previous example, one defines $z'$
and $z''$ similar to (\ref{dh.30}) and (\ref{dh.52}). Again
similar to the previous example, there is a relaxation time
corresponding to the fast phase, where from $z'$ and $z''$ at
least one is unimodular, which is obtained from (\ref{dh.39}),
with $|K|$ being the larger between $|K'|$ and $|K''|$. In order
that the system be in the slow phase, there should be a solution
where both $z'$ and $z''$ are nonunimodular. As for any value of
$E$, the product of the eigenvalues of $\tilde D_i$ is one, one
can take $z'$ and $z''$ so that both have moduli larger than one.
One then arrives at
\begin{equation}\label{dh.58}
W\,\begin{pmatrix}z''^{-1}&-1 \\ & \\
1& -z''\end{pmatrix}\,\tilde D_{L-L''+\mu\;L'-\mu}
\,\begin{pmatrix}z'^{-1}&-1 \\ & \\
1& -z'\end{pmatrix}\,V=0,
\end{equation}
or
\begin{equation}\label{dh.59}
\left[W\,\begin{pmatrix}z''^{-1}\\
\\1\end{pmatrix}\right]\,\left[\begin{pmatrix} 1 & -z''\end{pmatrix}\,\tilde
D_{L-L''+\mu\;L'-\mu} \,\begin{pmatrix}1\\
\\z'\end{pmatrix}\right]\,
\left[\begin{pmatrix}-z'^{-1} &1\end{pmatrix}\,V\right]=0.
\end{equation}
Putting the first or the third factor equal to zero, a result
similar to that of the previous example is obtained. But there may
be another solution which is obtained by letting the second factor
vanish:
\begin{equation}\label{dh.60}
\begin{pmatrix} 1 & -z''\end{pmatrix}\,\tilde
D_{L-L''+\mu\;L'-\mu} \,\begin{pmatrix}z'^{-1}\\
\\1\end{pmatrix}=0.
\end{equation}
This new solution (if it exists) can change the relaxation time.
If the relaxation time corresponding to the solution to
(\ref{dh.60}) is larger, then this relaxation time is the
relaxation time in the slow phase. So, the introduction a finite
part in the lattice can change the relaxation time and hence the
coexistence surface. In fact, qualitatively this finite part plays
the role of new boundaries introduced in the lattice.

\item[\textbf{4}] Alternating Coupling constant $K_\alpha=(-1)^{\alpha-\mu}\, K$

Defining
\begin{align}\label{dh.61}
\tilde D_\mathrm{o}:=\begin{pmatrix}0&1 \\ & \\ 1&
-(2+E)\,\coth(2\,K)\end{pmatrix}\nonumber\\ \tilde
D_\mathrm{e}:=\begin{pmatrix}0&1 \\ &
\\ 1& (2+E)\,\coth(2\,K)\end{pmatrix},
\end{align}
it is seen that (\ref{dh.26}) becomes
\begin{equation}\label{dh.62}
W\,\tilde D^{1-r}_\mathrm{o}\,(\tilde D_\mathrm{e}\,\tilde
D_\mathrm{o})^{(L-2+r)/2}\,V=0,
\end{equation}
where $r$ is $0$ ($1$) when $L$ is even (odd). Defining $z$
through
\begin{equation}\label{dh.63}
E=-2+(z+z^{-1})\,\tanh(2\,K),
\end{equation}
it is then seen that
\begin{equation}\label{dh.64}
\tilde D_\mathrm{e}\,\tilde D_\mathrm{o}=\begin{pmatrix}1&-(z+z^{-1})\\
&
\\(z+z^{-1})&1-(z+z^{-1})^2\end{pmatrix},
\end{equation}
showing that the eigenvalues of $(\tilde D_\mathrm{e}\,\tilde
D_\mathrm{o})$ are $(-z^2)$ and $(-z^{-2})$. So,
\begin{equation}\label{dh.65}
(\tilde D_\mathrm{e}\,\tilde D_\mathrm{o})^q=
\frac{(-1)^q}{z^{-1}-z}\,\begin{pmatrix}1&1\\ & \\
z&z^{-1}\end{pmatrix}\,
\begin{pmatrix}z^{2\,q}&0 \\ &\\0&z^{-2\,q}\end{pmatrix}\,
\begin{pmatrix}z^{-1}&-1\\&\\-z&1\end{pmatrix}.
\end{equation}
Again (\ref{dh.63}) shows that $\tau_0$ is the relaxation time of
the system (the system is in the fast phase), if all of the values
of $z$ are unimodular. To have a larger relaxation time (the
system be in the slow phase), there should be a value of $z$ with
modulus greater than 1. In fact, there should be a value of $z$
satisfying (\ref{dh.43}). In that case, in the thermodynamic limit
($q\to\infty$) one arrives from (\ref{dh.62}) to
\begin{equation}\label{dh.66}
\left[W\,\tilde
D^{1-r}_\mathrm{o}\,\begin{pmatrix}z^{-1}\\&\\1\end{pmatrix}\right]\,
\left[\,\begin{pmatrix}z^{-1}&-1\end{pmatrix}\,V\right]=0,
\end{equation}
or
\begin{equation}\label{dh.67}
\left[W\,\begin{pmatrix}(-1)^{1-r}\,z^{-1}\\&\\1\end{pmatrix}\right]\,
\left[\,\begin{pmatrix}z^{-1}&-1\end{pmatrix}\,V\right]=0.
\end{equation}
It is seen that for $r=1$ (odd $L$), this is exactly
(\ref{dh.42}). For $r=0$ (even $L$), the second factor in the
left-hand side of (\ref{dh.66}) is the same as the second factor
in (\ref{dh.42}), while the first factor is the first factor in
(\ref{dh.42}) with $z$ and $K$ replaced by $(-z)$ and $(-K)$,
respectively. So the coexistence surfaces (\ref{dh.44}) and
(\ref{dh.45}) are recovered here as well, while the coexistence
surfaces corresponding to $h_i$'s are the same as those obtained
for the uniform lattice if $L$ is odd, and the same with $K$
replaced by $(-K)$ if $L$ is even.

One could also arrive at this result by making a correspondence between
this case (alternating coupling constants) and case 1 (uniform coupling constants).
The analogy goes as the following.
\begin{align}\label{dh.68}
\tilde x_j&:=(-1)^{[j/2]}\,x_j,\nonumber\\
\tilde K_\alpha&:=(-1)^{\alpha-\mu}\,K_\alpha,\nonumber\\
(\tilde g_1,\tilde g_2,\tilde g_3,\tilde g_4)&:=(g_1,g_2,g_3,g_4),\nonumber\\
(\tilde h_1,\tilde h_2,\tilde h_3,\tilde h_4)&:=\begin{cases}(h_1,h_2,h_3,h_4),&\hbox{for odd $L$}\\
                                         (h_2,h_1,h_4,h_3),&\hbox{for even $L$}
                           \end{cases},
\end{align}
where $[y]$ means the largest integer not greater than $y$.
It is seen that in this way, if the untilded varibles and constants satisfy
(\ref{dh.10}) till (\ref{dh.12}), the tilded varibles and constants satisfy
(\ref{dh.10}) till (\ref{dh.12}) as well, and if $K_\alpha$ is alternating,
$\tilde K_\alpha$ is obviously uniform. The meaning of this correspondence
between $\tilde x_j$'s and $x_j$'s is that, for example, $x_0$ and $x_1$ are kept fixed,
while $x_2$ and $x_3$ have been changed to $(-x_2)$ and $(-x_3)$, respectively,
and $K_{1/2}$ and $K_{5/2}$ have been kept fixed, while $K_{3/2}$ has been changed to
$(-K_{3/2})$. Obviously the expressions $(K_{1/2}\,x_0\,x_1)$, $(K_{3/2}\,x_1\,x_2)$,
$(K_{5/2}\,x_2\,x_3)$ have all remained intact, while the coupling is now uniform, if
it has been alternating. If $L$ is odd, no change in the boundary rates is required
and the coexistence curves would be exactly the same as those found in the case of
uniform coupling. If $L$ is even, that part of the coexistence curve which arises from
the $g$'s is still intact, as $g$'s have remained intact. But the part due to $h$'s
is changed. And the change comes from the fact that $(h_2-h_1)$ has been changed to
$(h_1-h_2)$, while $(h_2+h_3)$ has remained intact. One can see that in the equations
of the coexistence curves involving $h$'s ((\ref{dh.44}) and (\ref{dh.45}) but with $g$'s
replaced with $h$'s), this change is equivalent to a change of $z$ to $(-z)$ and $K$ to
$(-K)$.
\end{itemize}

\section{Concluding remarks}
A one dimensional Ising model with nonuniform coupling constants
on a lattice with boundaries was studied. Detailed balance was
used to obtain the evolution of this model. The relaxation of the
resulted system towards its stationary solution was studied. A
general formulation was obtained to determine the relaxation times
and possible dynamical phase transitions, where the system moves
from the fast phase (with the relaxation time independent of
boundaries) to the slow phase (with the relaxation time depending
on boundaries). Some special examples were studied in more detail,
including cases where the lattice consists of two infinite
homogeneous parts and one middle (possibly nonhomogeneous part),
where it was shown that this middle part can induce further
dynamical phase transitions, behaving like an effective boundary.

Part of the results obtained here is similar to those corresponding
to the case of uniform coupling constants. The similarity is that
the system is in the fast phase when the boundary couplings are
high enough so that it is the bulk reaction rates that determine
the relaxation, and goes to the slow phase when the boundary reactions
are less than some critical value. Another similarity is that each
boundary behaves essentially independent to the other boundary. When
the system consists of large parts which are essentially uniform,
the new feature in the relaxation is that the system behaves as though
it is consists of independent systems each having two boundaries. So
for such systems new coexistence curves arise. There are of course
cases where no large uniform blocks are present. Then the only
similarity with the uniform case is that increasing the boundary
rates makes them unimportant in the relaxation time.

\section{Appendix}
\subsection{notation for a lattice}
A directional lattice is a collection of sites and directional links, with a
relation between links and sites. A link $\mathbf{l}$ has a negative boundary
(denoted by $\partial^-\mathbf{l}$), and a positive boundary (denoted by
$\partial^+\mathbf{l}$). The relation of the site $i$ to the link
$\mathbf{l}$ is that either $i$ is $\partial^-\mathbf{l}$, or
$i$ is $\partial^+\mathbf{l}$, or $i$ is not in $\mathbf{l}$. To each
site $i$ there corresponds a set of outgoing links $S^+(i)$, and a set of
incoming links $S^-(i)$:
\begin{align}\label{dh.69}
S^+(i)&:=\{\mathbf{l}\;|\;\partial^-\mathbf{l}=i\},\nonumber\\
S^-(i)&:=\{\mathbf{l}\;|\;\partial^+\mathbf{l}=i\}.
\end{align}
In a one dimensional (connected) lattice these relations become simple. The lattice sites
can be denoted by integers from a subset $Q$ of the integers, with the property that
if an integer is between two members of $Q$, it is a member of $Q$ itself. $Q$ can be
bounded from either above or below. If it is bounded from above, then the lattice has
a positive boundary, denoted by $\partial^+Q$, which is the largest member of $Q$.
Similarly the smallest member of $Q$ (if it exists) is the negative boundary of
$Q$ which is denoted by $\partial^-Q$. $S^+(i)$ consists of exactly one link, unless
$i$ is $\partial^+Q$, in which case $S^+(i)$ is empty. Similarly, $S^-(i)$ consists of
exactly one link, unless $i$ is $\partial^-Q$, in which case $S^-(i)$ is empty. So one can
denote the outgoing and incoming links corresponding to the site $i$, simply through
\begin{equation}\label{dh.70}
i\pm\mu:=S^\pm(i),
\end{equation}
from which one also has
\begin{equation}\label{dh.71}
\partial^\pm\alpha=\alpha\pm\mu.
\end{equation}
If the lattice is closed and has $L$ sites, one can still use the above notation,
provided an equivalence between $b$ and $(b+L)$ is assumed, where $b$ can be a site
(an integer) or a link (an integer plus half).
\subsection{the relaxation time}
Denoting the deviation from the static configuration of the system by $x$,
it is seen that $x$ satisfies
\begin{equation}\label{dh.72}
\dot y=h\,y,
\end{equation}
which is nothing but the closed form of (\ref{dh.11}). To solve this,
one expresses $y$ as a linear combination of the (generalized) eigenvectors
of $h$:
\begin{equation}\label{dh.73}
y(t)=\sum_\lambda c_\lambda(t) x_\lambda,
\end{equation}
where $x_\lambda$ is the (generalized) eigenvector of $h$ corresponding to the eigenvalue
$E_\lambda$. If all of the generalized eigenvectors are eigenvectors, evolution equations
for $c_\lambda$'s decouple as
\begin{equation}\label{dh.74}
\dot c_\lambda=E_\lambda\,c_\lambda,
\end{equation}
resulting in
\begin{equation}\label{dh.75}
c_\lambda(t)=c_\lambda(0)\,\exp(E_\lambda\,t).
\end{equation}
So one arrives at
\begin{equation}\label{dh.76}
y(t)=\sum_\lambda\,c_\lambda(0)\,\exp(E_\lambda\,t)\,x_\lambda.
\end{equation}
Initial condition determines $c_\lambda(0)$'s. It is seen that for large times
the leading term of the right hand side of (\ref{dh.76}) is that term which
corresponds to the largest real part of $E_\lambda$, unless the initial condition
is fine tuned so that $c_\lambda(0)$ vanishes for that $\lambda$. So the relaxation
time is
\begin{equation}\label{dh.77}
\tau=-\frac{1}{\max[\mathrm{Re}(E_\lambda)]}.
\end{equation}
In the thermodynamic limit, it may happen that the maximum of $\mathrm{Re}(E_\lambda)$
tends to zero, so that the relaxation time goes to infinity, meaning that the
relaxation is no longer exponential but say power law. But it may also happen
that this is not the case, and the maximum of $\mathrm{Re}(E_\lambda)$ tends to a
negative number, in which case the relaxation time remains finite even in the
thermodynamic limit.

\textbf{Acknowledgement}:  This work was partially supported by
the research council of the Alzahra University.

\end{document}